\documentclass[aps,prl,twocolumn]{revtex4-2}

\usepackage{graphicx,color}
\usepackage{bm,hyperref,amsmath}
\usepackage{amssymb}
\begin{document}

\title{Dynamical phase and quantum heat at fractional frequencies}
\author{George Thomas}
\affiliation{QTF Centre of Excellence, Department of Applied Physics, Aalto University, P.O. Box 15100, FI-00076 Aalto, Finland}
\affiliation{VTT Technical Research Centre of Finland Ltd, Tietotie 3, 02150 Espoo, Finland}
\author{Jukka P. Pekola}
\affiliation{QTF Centre of Excellence, Department of Applied Physics, Aalto University, P.O. Box 15100, FI-00076 Aalto, Finland}

\begin{abstract}
We demonstrate a genuine quantum feature of heat: the power emitted by a qubit (quantum two-level system) into a reservoir under continuous driving shows peaks as a function of frequency $f$. These resonant features appear due to the accumulation of the dynamical phase during the driving. The position of the $n$th maximum is given by $f=f_{\rm M}/n$, where $f_{\rm M}$ is the mean frequency of the qubit in the cycle, and their positions are independent of the form of the drive and the number of heat baths attached, and even the presence or absence of spectral filtering. We show that the waveform of the drive determines the intensity of the peaks, differently for odd and even resonances.  
This quantum heat is expected to play a crucial role in the performance of driven thermal devices such as quantum heat engines and refrigerators. We also show that by optimizing the cycle protocol, we recover the favorable classical limit in fast driven systems without the use of counter-diabatic drive protocols  and we demonstrate an entropy preserving non-unitary process. We propose that this non-trivial quantum heat can be detected by observing the steady-state
power absorbed by a resistor acting as a bolometer attached to a driven superconducting qubit.
\end{abstract}

\maketitle
\textit{ Introduction:}
Quantum heat transport \cite{Hanggi,Dubi,Alberto,RMP} and  quantum heat engines and refrigerators \cite{Sai2016,Pekola2016,Deffner2019,Pekola2019} attract currently attention because of their role in thermodynamics in quantum domain, and their applicability in such areas as heat management
in quantum circuits, qubit resetting for quantum computational tasks \cite{Tan2017}, and quantum error correction \cite{Danageozian}. In this context, whether quantum coherence is detrimental or advantageous for the performance
of the thermal machines is currently a much debated topic \cite{Pekola2019, Kosloff2002,Scully2003,Scully2011,Holubec2018,Brandner2016, Streltsov2017,Du2018, Kilgour2018, Holubec2018, Camati2019, Dann2020, Hammam2021}.
Moreover, non-classical nature of heat originates from the coherence terms of the density matrix when expressed in the eigenbasis of the Hamiltonian
\cite{Elouard2017,Elouard2020}.
Therefore understanding the role of coherence is important: in this letter we demonstrate the role of coherence in the emitted power by a qubit under continuous-wave high-frequency driving. In this driven open quantum system,
power versus driving frequency has resonances at well-defined frequencies, due to the accumulation of the dynamical phase of the qubit \cite{Pekola2016, Grifoni1998, Weiss2016, Russomanno2012, Russomanno2017, Silveri2015, Shevchenko2012, Utkarsh2021, Andrew2022}.
In fact these peaks are general and appear irrespective of whether the qubit is coupled to the bath(s) with or without spectral filtering, and of the number of baths attached. For simplicity to obtain transparent results, we first analyse a driven qubit coupled to a single bath and the form of the drive is approximated by a square wave.
Here we give a theoretical explanation of the characteristics and origin of these resonances. With a single qubit, the positions of the 
peaks align with  $f_{\rm M}/n$ dependence, where $f_{\rm M}$ is the mean frequency of the qubit in the cycle and $n$ is an integer.
\begin{figure} [h!]
	\begin{center}
		\includegraphics[scale=1.02]{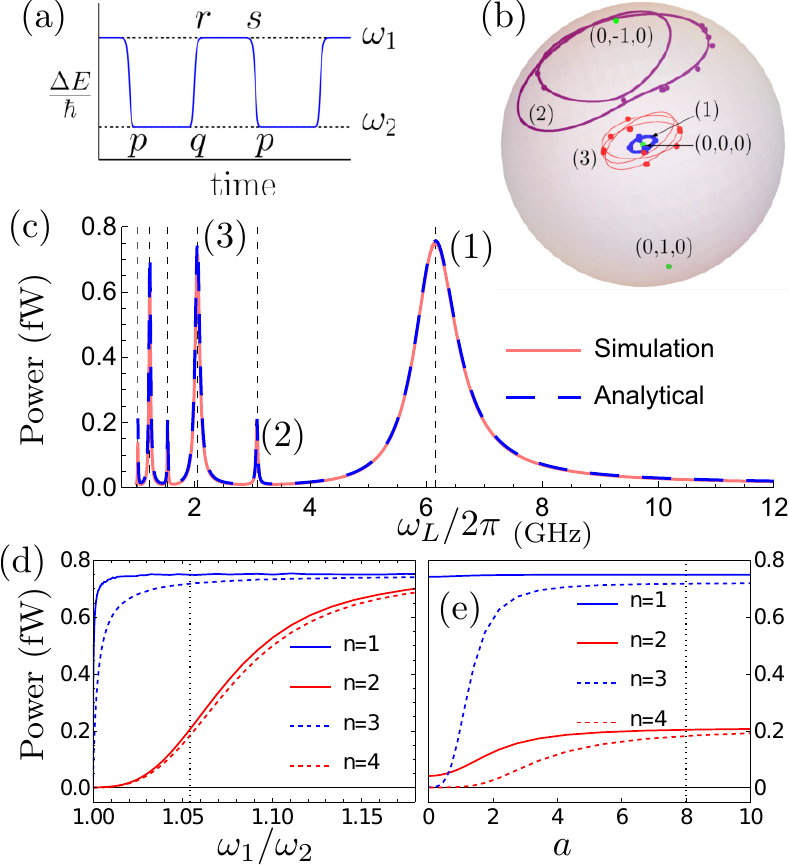}
		\caption{(a), A pictorial representation of the driving protocol for $a=8$. 
			(b), Bloch sphere representation of the trajectories at the driving frequency $f_{L,1}$, $f_{L,2}$ and $f_{L,3}$. Dots on the trajectories are obtained using Eqs. (\ref{ramps}). (c), Power dissipated in the bath, where the red curve represents simulation using Eqs. (\ref{master_eq}), (\ref{dissipator}) and (\ref{exact_power}) and blue dashed curve is obtained 
			analytically from Eqs. (\ref{ramps}) and (\ref{approx_power}). 
			The dashed vertical lines are at
			$f_{L, n} $ with $n=1,2,...,6$ from right to left. 
			The peaks denoted with (1), (2) and (3) in (c) correspond to the trajectories shown in (b).
			Here we have ignored pure dephasing effects and taken typical parameters for a superconducting qubit $\omega_0/2\pi=6$ GHz, $g/2\pi=1$ GHz, $T=70$ mK, and $\kappa= 0.01$. Power at the maxima of peaks of different order, $n=1,2,3,4$, as functions of $\omega_1/\omega_2$ in (d), and of $a$ in (e). The vertical dashed lines in (d) and (e) correspond to the parameter values in panel (c).}
		\label{peaks}
	\end{center}
\end{figure}

\textit{Model:}
First we consider a driven qubit coupled to a heat bath
at temperature $T$.
The  Hamiltonian of the qubit is given as
\begin{equation}
H(t)=\frac{\hbar g \Omega(t)}{2}\sigma_z+ \frac{\hbar \omega_0}{2}\sigma_x,
 \label{ham_Q}
\end{equation}
 where $\sigma_z$ and $\sigma_x$ are the Pauli matrices and $\omega_0$ is the minimum transition frequency of the qubit. Here we have the driving protocol  (see Fig. \ref{peaks} (a))
\begin{equation}
\Omega(t)=1+\frac{\tanh{[a\cos{(\omega_L t)}]}}{\tanh{a}},
\label{drive}
\end{equation}
where $\omega_L$ is the frequency, $g$ is the amplitude of the drive and $a$ is a real parameter. When $a\rightarrow0$, $\Omega(t)=1+\cos{(\omega_L t)}$ and when $a\rightarrow\infty$, $\Omega(t)$ is a square wave of  unit amplitude. 
 From Eq. (\ref{ham_Q}), we get the energy gap between ground and excited states of the qubit at any instant of time as 
 \begin{equation}
\Delta E (t)=\hbar \sqrt{g^2 \Omega(t)^2+\omega_0^2}.
\end{equation}
For $\Omega(t)=2$, and $\Omega(t)=0$, we get maximum  and minimum energy level spacings $\Delta E_1$ and $\Delta E_2$, respectively. Corresponding transition frequencies
are denoted as $\omega_1=\Delta E_1/\hbar$ and $\omega_2=\Delta E_2/\hbar$.
We consider a weak coupling between the system and the bath. 
The density matrix $\rho$ of the 
system undergoes a non-unitary evolution as  \cite{Breuer2002}
\begin{equation}
 \frac{d\rho}{d t}=-\frac{i}{\hbar}[H(t),\rho]+{\cal L}\rho+{\cal L}^{\phi}\rho,
 \label{master_eq}
\end{equation}
where the dissipator is given by
\begin{eqnarray}
 {\cal L}\rho&=&\Gamma^{\downarrow}(\sigma^{-}\rho \sigma^{+}-1/2\{\sigma^{-} \sigma^{+},\rho\})
 \nonumber\\
 &+&\Gamma^{\uparrow}(\sigma^{+}\rho \sigma^{-}-1/2\{\sigma^{+} \sigma^{-},\rho\}),
 \label{dissipator}
\end{eqnarray}
and pure dephasing by ${\cal L}^{\phi}\rho=\Gamma^{\phi}(\sigma_z\rho \sigma_z-\rho)$.
Here $\sigma^{+}$ and $\sigma^{-}$
 are raising and lowering operators, respectively.
For the case of coupling to the bath in $\sigma_z$ direction, the transition rates are \cite{Pekola2016,Pekola2019}
\begin{equation}
\Gamma^{\downarrow}=\kappa \frac{\omega_0^2}{\omega_0^2+g^2 \Omega(t)^2} \frac{\Delta E}{\hbar} (N(\Delta E)+1).
\label{Gamma}
\end{equation}
 Here $\kappa$ is dimensionless coupling parameter, $N(\Delta E)=1/(\exp{ (\Delta E/k_B T) }-1)$,
 $\Gamma^{\uparrow}= \exp{ (-\Delta E/k_B T) }\Gamma^{\downarrow}$ due to detailed balance,  and from zero frequency  noise spectrum,  we get the pure dephasing rate as \cite{Vavilov2014} $\Gamma^{\phi}=\kappa(\omega_0^2/[g\Omega(t)]^2+1)^{-1} k_B T/\hbar$.

\textit{Origin of peaks in a fast driven system:}
Here we analyse a case where $\omega_0 \gg g$ in the fast driven regime $\omega_L\gg\Gamma^{\downarrow},\Gamma^{\uparrow}$. We can then ignore the effects due to pure dephasing as $g^2 \Omega(t)^2\ll \omega_0^2$. For sufficiently large $a$, the drive is close to a  square-wave as shown in Fig.~\ref{peaks}(a). Thus for
 half of the period $\delta t=\pi/\omega_L$, the energy of the system is $\Delta E =\Delta E_2$, and for the other half, it is $\Delta E=\Delta E_1$ and the transitions between these legs during the drive can be approximated as sudden processes. Thus for $a\rightarrow \infty$, baths are acting on the system only in the branches $\Delta E=\Delta E_2$ and $\Delta E=\Delta E_1$, and the relaxation rates (see Eq. (\ref{Gamma})) at these branches are denoted as $\Gamma^{\downarrow}_2$
and $\Gamma^{\downarrow}_1$, respectively.
Thus we can identify four steps during the drive: $p\rightarrow q$,  $r\rightarrow s$ are the thermalization steps and
 $q\rightarrow r$,  $s\rightarrow p$ represent the sudden changes as shown in  Fig.~\ref{peaks}(a).
 In the fast driven system, since the Hamiltonians at two 
different instances are not commuting $[H(t'),H(t'')]\neq0$, 
the coherences created dissipate heat to the baths due to the work performed on the system during the driving \cite{Kosloff2002}. 
From Eqs. (\ref{ham_Q}) and (\ref{master_eq}), we get the power dissipated in the bath in a cycle as
\begin{eqnarray}
 P=\frac{\omega_L}{2 \pi}\int_0^{2\pi/\omega_L}{\rm Tr} [H(t){\cal L}\rho]\, dt.
 \label{exact_power}
 \end{eqnarray}
This expression coincides with that for the heat associated to open quantum evolution in previous literature \cite{Alicki,Solinas,Esposito}.
 For $a\gg 1$, we can approximate
 \begin{eqnarray}
 P=\frac{\omega_L}{2 \pi}\left[\Delta E_2({\mathcal D}_q-{\mathcal D}_p)+\Delta E_1({\mathcal D}_s-{\mathcal D}_r)\right],
 \label{approx_power}
\end{eqnarray}
where ${\mathcal D}_i= 1/2-\rho_{ee,i}$, with $\rho_{ee,i}$ the occupational probability of the excited state at the $i^{\rm th}$ position of the cycle.
Interestingly, the power versus driving frequency shows maxima as depicted in Fig~\ref{peaks}(c). To explain this, we consider the dynamical phase $\varphi$ accumulated in the off-diagonal (coherence) terms of
 the density matrix which is in the form $\exp{(-i\varphi)}$, where
 \begin{equation}
  \varphi=\frac{1}{\hbar}\int_0^{2 \pi/\omega_L}\Delta E(t) dt= (\Delta E_1+\Delta E_2 )\frac{\pi}{\omega_L}, \label{eq:Phi}
 \end{equation}
valid for any value of $a$.
When the accumulated phase  $\varphi=2 n \pi$, where $n$ is an integer, the system makes $n$ complete rotations in a Bloch sphere.
Equating $(\Delta E_1+\Delta E_2 )\pi/\hbar\omega_L =2 n \pi$,
  we get the condition of these frequencies
 \begin{equation}
  f_{L,n}=\frac{f_{\rm M}}{ n},
  \label{position}
 \end{equation}
  where $f_{\rm M}=(\Delta E_1+\Delta E_2)/2\,h$.
For each $n$, the system undergoes a trajectory with $n$ closed  loops (See Fig.\ref{peaks}(b)). 
  An immediate consequence of Eq. (\ref{eq:Phi}) is that 
 the positions of the peaks are robust irrespective of the waveform of the drive. 
 
 Equation (\ref{position}) can also be achieved by considering the time-evolution of the density matrix in a fast driven system as we will now demonstrate.
We find an analytical expression for the density matrix and the dissipated power in the limit $a\rightarrow \infty$, \emph{i.e.} for the abrupt changes of the energy of the qubit.  
 We define  ${\mathcal R}_i$ and  ${\mathcal I}_i$ as the real and imaginary  parts of the off-diagonal terms of the density matrix of the system, respectively. These are defined in the instantaneous eigenstates of the Hamiltonian. Their evolution can be obtained by applying the sudden approximation of quantum mechanics  in legs $q\rightarrow r$ and $s\rightarrow p$ and relaxation  in legs $p\rightarrow q$ and $r\rightarrow s$.
 In steady state they  evolve  as
\begin{eqnarray} 
&&{\mathcal D}_q = {\mathcal D}_p+[\Gamma^\downarrow_2-\Gamma^\Sigma_2 ({\mathcal D}_p+1/2)]\delta t_2\nonumber\\
&&
\,\,{\mathcal R}_q = \left[{\mathcal R}_p \cos{(\omega_2 \delta t_2)}-{\mathcal I}_p \sin{(\omega_2 \delta t_2)} \right] (1-\frac{1}{2}\Gamma^\Sigma_2 \delta t_2)\nonumber\\
&&\,\,{\mathcal I}_q= \left[{\mathcal I}_p \cos{(\omega_2 \delta t_2)}+{\mathcal R}_p \sin{(\omega_2 \delta t_2)}\right](1-\frac{1}{2}\Gamma^\Sigma_2 \delta t_2)={\mathcal I}_r \nonumber \\
&&{\mathcal D}_r = \sqrt{1-\eta^2}{\mathcal D}_q -\eta{\mathcal R}_q
, {\mathcal R}_r
=  \sqrt{1-\eta^2}{\mathcal R}_q+\eta{\mathcal D}_q,
\nonumber \\
&&{\mathcal D}_s= {\mathcal D}_r+[\Gamma^\downarrow_1-\Gamma^\Sigma_1 ({\mathcal D}_r+1/2)]\delta t_1\nonumber\\
&&
\,\,{\mathcal R}_s = [\cos{(\omega_1 \delta t_1)}\mathcal R}_r  {-{\mathcal I}_q \sin{(\omega_1 \delta t_1)}](1-\frac{1}{2}\Gamma^\Sigma_1 \delta t_1)\nonumber \\
&&\,\,{\mathcal I}_s = [\cos{(\omega_1 \delta t_1)} {\mathcal I}_q +{\mathcal R}_r \sin{(\omega_1 \delta t_1)}](1-\frac{1}{2}\Gamma^\Sigma_1 \delta t_1)={\mathcal I}_p\nonumber \\
&&{\mathcal D}_p =  \sqrt{1-\eta^2}{\mathcal D}_s+ \eta{\mathcal R}_s,\,\,
{\mathcal R}_p =   \sqrt{1-\eta^2}{\mathcal R}_s-\eta{\mathcal D}_s.
\label{ramps}
\end{eqnarray}
 Here we consider $\Gamma^{\Sigma}_1 \delta t_1\ll1$ and $\Gamma^{\Sigma}_2\delta t_2\ll1$, where $\Gamma^{\Sigma}_{1(2)} =\Gamma^{\uparrow}_{1(2)}+\Gamma^{\downarrow}_{1(2)}$ and $\delta t_1$ and $\delta t_2$ are the durations of the legs $p\rightarrow q$ and $r\rightarrow s$, respectively such that $\omega_L=2\pi/(\delta t_1+\delta t_2)$, and $\eta=\sqrt{1-\omega_2^2/\omega_1^2}$.  For the symmetric case $\delta t_1=\delta t_2$, Eqs. (\ref{ramps}) represent the evolution of the system corresponding to the driving
protocol in Eqs. (\ref{ham_Q}) and (\ref{drive}) with $a\rightarrow \infty$, and below we show that it yields identitical results with the full numerical solution of the master equation (Eq. (\ref{master_eq})).
 Naturally the maximum occupation probability of the excited state at the point $p$ versus driving frequency appears at the same position as that of power maxima shown in Fig. \ref{peaks} (c).
From Eqs. (\ref{ramps}), and considering $\Gamma^\downarrow_1=\Gamma^\Sigma_1=0$, $\delta t_2=\delta t_1=\delta t$ and $\Gamma^\downarrow_2\delta t\ll 1$, we get
\begin{eqnarray}
&& P=\frac{\Delta E_2 (2\Gamma^\downarrow_2-\Gamma^\Sigma_2)[1-\cos( \omega_1 \delta t)](\omega_1^2-\omega_2^2)
 }{2\left(4\omega_1^2-(\omega_1+\omega_2)^2\cos[ (\omega_2+\omega_1) \delta t]-{\mathcal K}\right)},
 \label{condition}\\
 && {\rm and} \;\;  \rho_{ee,p}=\frac{1}{2}-\frac{(2\Gamma^\downarrow_2-\Gamma^\Sigma_2)}{\Gamma^\Sigma_2}\nonumber\\
 &&\times \frac{4\left[\omega_1 \cos\frac{\omega_1 \delta t}{2}\sin\frac{\omega_2 \delta t}{2}+\omega_2\cos\frac{\omega_2 \delta t}{2}\sin\frac{\omega_1 \delta t}{2}\right]^2}{\left(4\omega_1^2-(\omega_1+\omega_2)^2\cos[ (\omega_2+\omega_1) \delta t]-{\mathcal K}\right)},
 \label{conditionPg}
\end{eqnarray}
with
$
 {\mathcal K}=
  2 (\omega_1^2-\omega_2^2)\cos [\omega_2\delta t] +(\omega_1-\omega_2)^2\cos [(\omega_1-\omega_2) \delta t].\nonumber
$
The analytical results above are valid for small amplitude ($\omega_1\approx \omega_2$), square-wave driving ($a\rightarrow \infty$) of the qubit. It is, however, illustrative to look at different driving protocols to understand the various peaks and their origin. In Fig. \ref{peaks} (d), the amplitudes of the various peaks are presented against $\omega_1/\omega_2$. We see that the odd $n$ peaks assume their asymptotic value already with weak driving, whereas the even $n$ peaks grow more gradually. This is in line with the results in panels (b) and (c), where the even peaks behave very differently (on Bloch sphere, and in terms of the height and width of the peaks) with respect to odd $n$ peaks. In Fig. \ref{peaks} (e) the same quantities are plotted now as functions of $a$. Only the $n=1$ peak survives all the way to $a\rightarrow 0$, i.e. for a sinusoidal drive, whereas the rest of the peaks arise only for non-vanishing $a$. This observation suggests that the magnitude of the $n$th peak is determined by the corresponding Fourier component of the drive.

When $\omega_1\rightarrow\omega_2$,  we get from Eqs. (\ref{condition}) and (\ref{conditionPg}) maximum value for $P$ and $\rho_{ee,p}$ for $\delta t =2 n\pi/(\omega_2+\omega_1)$,  which corresponds to Eq.~(\ref{position}) and to the peaks obtained in the power. The condition used in Eqs. (\ref{condition}) and (\ref{conditionPg}), $\Gamma^\downarrow_1=\Gamma^\Sigma_1=0$, can be easily achieved by using spectral filter with sufficiently high quality factor \cite{Alberto}. Interestingly, when $\delta t=2 n \pi/\omega_1$, $P=0$, and $\rho_{ee,p}=\Gamma^\uparrow_2/\Gamma^\Sigma_2$, which is the classical limit \cite{Pekola2016}. In this case, ${\mathcal R}_r= {\mathcal R}_s$, 
${\mathcal I}_r= {\mathcal I}_s$ and ${\mathcal D}_r= {\mathcal D}_s$ due to which the coherence created during the ramp
$q\rightarrow r$ is annihilated in the ramp $s\rightarrow p$. Generally, transitionless quantum driving 
is achieved by counter-diabatic driving \cite{Demirplak2003,Berry2009}. Here, we achieve the classical limit with minimal power without such counter-diabatic driving but with suitable choice of the driving protocol.

Another implication of $\Gamma^\downarrow_1=\Gamma^\Sigma_1=0$ is that the non-unitary step where the system
is in contact with the heat bath should preserve the purity of the qubit. Consider $U_{ij}$ as the unitary process representing the ramp $i\rightarrow j$ where  $i,j=\{p,q,r,s\}$. From the cyclic process described in Fig.~\ref{peaks}(a), we have
\begin{equation}
 {\cal V}_{pq}\{ U_{sp}U_{rs} U_{qr}\rho_q U_{qr}^{\dagger}U_{rs}^{\dagger}U_{sp}^{\dagger}\}=\rho_q,
 \label{cyclic_cond}
\end{equation}
 where $\rho_q$ is the density matrix at the beginning of the ramp $q \rightarrow r$ and ${\cal V}_{pq}$ represents the map corresponding to non-unitary process in the branch $p\rightarrow q$.  Unitary processes preserve the von-Neumann entropy (purity) of the system.  To  achieve cyclicity  as shown in Eq. (\ref{cyclic_cond}), purity of the system in non-unitary branch $p\rightarrow q$ should also be preserved. This implies,
\begin{equation}
 {\cal D}_{i}^2+{\cal R}_{i}^2+{\cal I}_{i}^2={\cal D}_{j}^2+{\cal R}_{j}^2+{\cal I}_{j}^2.
\end{equation}
Under Lindblad  evolution, due to decoherence, we have $\sqrt{{\cal R}_{q}^2+{\cal I}_{q}^2}<\sqrt{{\cal R}_{p}^2+{\cal I}_{p}^2}$, which implies ${\cal D}_{q}>{\cal D}_{p}$ and thereby $P>0$ as expected. 
A possible application of this effect can be mitigation of the loss of quantum information in an open system with drive.

\textit{Bloch sphere dynamics:}
The driven system shows interesting trajectories in the Bloch sphere representation. The co-ordinates of the Bloch vector at a given instant
of time are ($\langle\sigma_x\rangle_t, \langle\sigma_y\rangle_t,\langle\sigma_z\rangle_t$)
where $\langle\sigma_i\rangle_t={\rm Tr} [\sigma_i \rho(t)]$ and $\rho(t)$ is obtained from Eq. (\ref{master_eq}) for $t\rightarrow\infty$.
Since we are depicting steady state cyclic processes, all the trajectories in the Bloch sphere for a complete cycle should be closed. Depending on the driving frequency or in other words, the position of the peak, the number of turns in the Bloch sphere trajectory equals $n$ (See  Fig. \ref{peaks} (b)). For driving frequencies away from $f_M/n$, the trajectories are close to the surface (not shown in Fig. \ref{peaks} (b)).
They move towards the centre of the Bloch sphere when the driving frequencies are close to $f_M/n$ (peaks). This is due to the fact that at driving frequencies near $f_{L,n}$, the off diagonal coherence terms are created which then dissipate to the bath. Or in other words, the driving increases the entropy which in turn increases the mixedness of the system. Thus trajectories reflect the amount of dissipation. 
The trajectory on the Bloch sphere can also be constructed from 
Eqs. (\ref{ramps}). At a given instant, the co-ordinates are 
$2({\cal R},{\cal I},{\cal D})$. 
As an example, we represent the state of the system obtained from Eqs. (\ref{ramps}) at a few instances as dots in
Fig. \ref{peaks}(b). This shows an excellent agreement between analytical solution from Eqs. (\ref{ramps}) and numerical simulations.

\textit{Experimental setup:}
The predicted quantum heat might be observable in a basic qubit based setup.
 \begin{figure}
\begin{center}
\includegraphics[width=6cm]{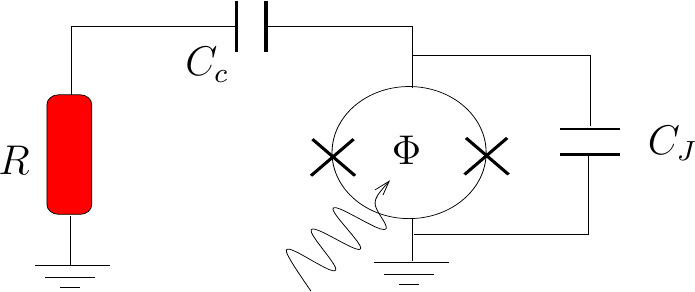}
\caption{A possible experimental setup consists of a superconducting qubit capacitively coupled to a normal metal resistor with resistance $R$.}
\label{experiment}
\end{center}
\end{figure}
For a transmon qubit \cite{Koch2007} with  $ \Delta E=\hbar \omega$, coupled  to a resistor as bath \cite{Nyquist} through a capacitor with capacitance $C_c$ (see Fig. \ref{experiment}), the transition rate
is given as
\begin{equation}
 \Gamma^{\downarrow}=\frac{\omega_0^2}{\omega_0^2+g^2 \Omega(t)^2} \frac{C_c^2}{C_{\Sigma}^2}\frac{\omega}{Q} (N(\Delta E)+1),
\end{equation}
where $C_J$ is the capacitance of each junction, $C_{\Sigma}=C_c+2C_J$,  quality factor of the junction $Q=\sqrt{L_J/C_J}/R=1/\omega C_J R$, and $L_J$ is the Josephson inductance which can be modulated with flux $\Phi$.  For $g \Omega(t)/\omega_0\ll 1$  and for low temperature $\hbar\omega>k_B T$, we get
\begin{equation}
 \Gamma^{\downarrow}\approx C_c^2C_J(C_J+C_c)^{-2}\omega^2 R C_J.
\end{equation}
For typical values of a transmon, $C_J=30$ fF, $C_c=8$ fF $\omega/2\pi= 6$ GHz, 
and $R=200$ $\Omega$, we  have $\Gamma^{\downarrow}/\omega\approx 0.01$ which represents weak coupling
and the proposed model is applicable.

\begin{figure}
\begin{center}
 \includegraphics[scale=0.94]{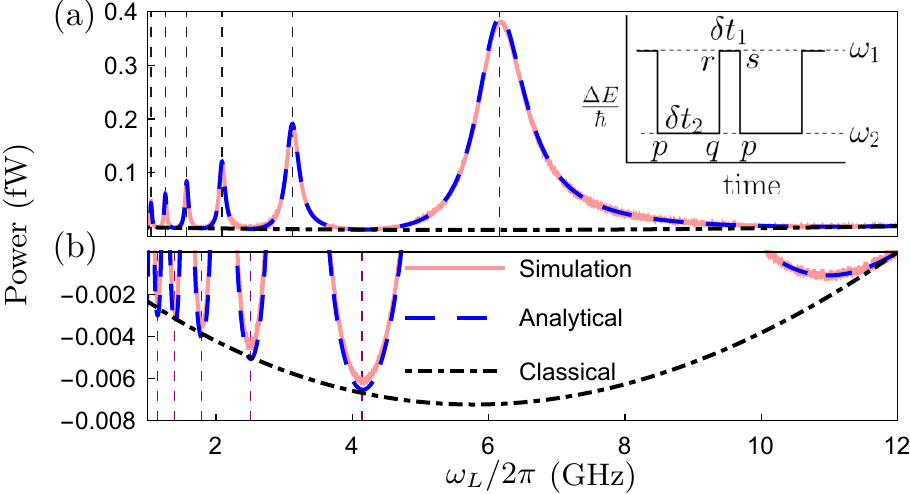}
\caption{(a), Power $P_2$ versus frequency. Panel (b) shows  the enlarged cooling regime. Inset shows the asymmetric square-wave driving protocol ($\delta t_2\neq\delta t_1$) used in (a) and (b).
The
dashed-blue and continous-red curves correspond to theoretical model from Eqs. (\ref{ramps}) and simulation, respectively.
The dot-dashed black curves indicate the classical limit with same energy level spacings and transition rates as in the quantum system. Vertical dashed lines in (a) are obtained from Eq. (\ref{position2}).
 Here $\omega_L$ is varied by changing $\delta t_1$. The vertical dashed lines in (b) correspond to $\delta t_1=2 n\pi/\omega_1$, where cooling is achieved. We take $\delta t_2=\pi/\omega_2$, $T_1=T_2=210$ mK, $\omega_0/2\pi=6$ GHz, $g/2\pi=1$ GHz, and $\kappa= 0.01$. }
\label{peaks_cooling}
\end{center}
\end{figure}

\textit{Cooling regime:}
We have seen that the classical limit or in other words  suppression of the coherence induced dissipation, can be achieved by considering $\Gamma^\downarrow_1=\Gamma^\Sigma_1=0$ and $\delta t=2 n \pi/\omega_1$.  
We can extend this approach for two baths coupled to the qubit via resonators to approach this limit. 
Such a setup will be useful in constructing quantum heat engines and refrigerators \cite{Andrew2022}.
The resonators (spectral filters) will help to couple the qubit to the  bath 1 with temperature $T_1$ when $\Delta E=\Delta E_1$
and  to the bath 2 at temperature $T_2$ when $\Delta E=\Delta E_2$, and allow almost unitary evolution
in between as discussed in Refs. \cite{Alberto,Pekola2019}. Now we find the power dissipated to baths 1 and 2 as $P_1=\left[\Delta E_1({\mathcal D}_s-{\mathcal D}_r)\right]\omega_L/2 \pi$ and 
$P_2=\left[\Delta E_2({\mathcal D}_q-{\mathcal D}_p)\right]\omega_L/2 \pi$, respectively.
The cooling regime is defined as $P_2<0$ and $P_1>0$ and can be achieved at high frequency by suitably choosing $\delta t_2$ and $\delta t_1$. 
If we consider the case $\delta t_2=\pi/ \omega_2$, we get cooling for $\delta t_1=2 n \pi/ \omega_1$
as shown in Fig. \ref{peaks_cooling}.
The transition rate due to the bath in the presence of resonators is given as \cite{Pekola2016}
\begin{equation}
\Gamma^{\downarrow}_r=\kappa \frac{\omega_0^2}{\omega_0^2+g^2 \tilde{\Omega}(t)^2} \frac{(N(\Delta E)+1)\Delta E/\hbar}{1+Q_r^2\left(\frac{\omega_r}{\omega}-\frac{\omega}{\omega_r}\right)^2},
\end{equation}
where $r=1,2$, $Q_r$ is the quality factor of the $r$th resonator, $\omega=\Delta E/\hbar$ and $\tilde{\Omega}(t)$ corresponds to
the drive shown in the inset of Fig. \ref{peaks_cooling}(a). 
  Moreover, the dynamical phase is $\varphi=\frac{1}{\hbar}\int_0^{2 \pi/\omega_L}\Delta E(t) dt= \pi+ \omega_1 (2 \pi/\omega_L-\pi/\omega_2)$. Invoking the condition $\varphi=2 n \pi$ as in Eq. (\ref{position}), we get the power maxima (see Fig. \ref{peaks_cooling})  at frequencies 
  \begin{equation}
  f_{L,n}^{\rm asy}=\frac{1}{2 \pi}\frac{2 \omega_2 \omega_1}{(2n-1) \omega_2+ \omega_1}.
  \label{position2}
 \end{equation} 
The validity of Eqs. (\ref{ramps}) is in the regime $\Gamma^\downarrow_1\delta t_1\ll 1$ and $\Gamma^\downarrow_2\delta t_2\ll 1$, because we consider only the linear terms  in $\delta t_1$ and $\delta t_2$ in the dissipators (see Eqs. (\ref{dissipator}) and (\ref{ramps})) in branches $p\rightarrow q$ and $r\rightarrow s$. Therefore we can see a slight mismatch between simulation  and analytical solution in Fig. \ref{peaks_cooling}.

Irrespective of the initial state of the  system, after sufficiently many periods of drive, the system reaches
a \textit {steady state cycle} with the same trajectory in the Bloch sphere for all the subsequent cycles.
At this point, an interesting question in the periodically driven open quantum system is: what would be the minimum relaxation rate required to make the evolution of the system cyclic \cite{Brander} and what would be the corresponding trajectory since for fully unitary (closed system) such steady state is not reached? This can be understood from Eqs. (\ref{ramps}) and (\ref{conditionPg}). For  $\Gamma^\downarrow_1=0$, the cyclic condition is satisfied for
any $\Gamma^\downarrow_2$ except for $\Gamma^\downarrow_2=0$. As $\Gamma^\downarrow_2\rightarrow0$, the cyclic trajectory of the system approaches the center of the Bloch sphere. So in Eq. (\ref{conditionPg}), we get $\rho_{ee,p}\rightarrow 1/2$ and thereby  ${\cal D}_p\rightarrow0$. Similar analysis
can be done for ${\cal R}$ and ${\cal I}$. As $\Gamma^\downarrow_2\rightarrow0$, the Bloch vector $2({\cal R},{\cal I},{\cal D})\rightarrow (0,0,0)$. Thus for an arbitrary initial state, away from the steady state trajectory,  the system takes infinitely many cycles  for 
 $\Gamma^\downarrow_2\rightarrow0$ to reach the steady state cycle. But for any non-vanishing $\Gamma^\downarrow_2$, a steady state
 cycle is eventually reached.

 To conclude, we have established the relation between the quantum heat in driven systems and the dynamical phase acquired during the drive.  We show that odd and even peaks of quantum heat have different origins and their intensity depends on the waveform of the drive. By manipulating the cycle protocol, one can approach the favorable classical limit without counter-diabatic drive and it is possible to preserve purity even in the presence of a heat bath. We discussed the trajectories traversed by the qubit on the Bloch sphere and the impact of dissipation on cyclicity. Our work can be extended to many interesting directions such as experimental verification of the 
proposed model, analysis of whether the system can outperform  the classical limit, and heat transport at other fractional frequencies in multilevel systems.

\begin{acknowledgments}
\textit{Acknowledgment}: We thank Bayan Karimi and Dmitry S. Golubev for useful discussions.
We acknowledge Academy of Finland grant 312057, the European Union’s Horizon 2020 research and innovation programme under the European Research Council (ERC) programme (grant number 742559) and Foundational Questions Institute Fund (FQXi) via Grant No. FQXi-IAF19-06. G.T also acknowledge the support by the Academy of Finland Flagship Programme, Photonics Research and Innovation (PREIN), decision number 320168.
\end{acknowledgments}

\end{document}